\documentclass[aps,prb Rapid Comm.,groupedaddress,superscriptaddress,reprint]{revtex4-1}
\usepackage{graphicx}
\usepackage{bpchem}
\usepackage{amssymb}

\begin{document}


\title{Thermal-history dependent magnetoelastic transition in \BPChem{(Mn,Fe)\_{2}(P,Si)}} 



\author{X. F. Miao}
\email{x.f.miao@tudelft.nl}
\affiliation{Fundamental Aspects of Materials and Energy, Faculty of Applied Sciences, Delft University of Technology, Mekelweg 15, 2629 JB Delft, The Netherlands}
\author{L. Caron}
\affiliation{Fundamental Aspects of Materials and Energy, Faculty of Applied Sciences, Delft University of Technology, Mekelweg 15, 2629 JB Delft, The Netherlands}
\author{Z. Gercsi}
\affiliation{Blackett Laboratory, Department of Physics, Imperial College London, London SW7 2AZ, United Kingdom}
\affiliation{CRANN and School of Physics, Trinity College Dublin, Ireland}
\author{{{A. Daoud-Aladine}}}
\affiliation{ISIS facility, Rutherford Appleton Laboratory, Chilton, Didcot, Oxfordshire, OX11 0QX, United Kingdom}
\author{N. H. van Dijk}
\affiliation{Fundamental Aspects of Materials and Energy, Faculty of Applied Sciences, Delft University of Technology, Mekelweg 15, 2629 JB Delft, The Netherlands}
\author{K. G. Sandeman}
\affiliation{Department of Physics, Brooklyn College, CUNY, 2900 Bedford Avenue, Brooklyn, New York 11210, USA}
\affiliation{CUNY Graduate Center, 365 Fifth Avenue, New York, New York 10016, USA}
\affiliation{Blackett Laboratory, Department of Physics, Imperial College London, London SW7 2AZ, United Kingdom}
\author{E. Br\"{u}ck}
\affiliation{Fundamental Aspects of Materials and Energy, Faculty of Applied Sciences, Delft University of Technology, Mekelweg 15, 2629 JB Delft, The Netherlands}

\date{\today}

\begin{abstract}

The thermal-history dependence of the magnetoelastic transition in \BPChem{(Mn,Fe)\_{2}(P,Si)} compounds has been investigated using high-resolution neutron diffraction. As-prepared samples display a large difference in paramagnetic-ferromagnetic (PM-FM) transition temperature compared to cycled samples. The initial metastable state transforms into a lower-energy stable state when the as-prepared sample crosses  the  PM-FM transition for the first time. This additional transformation is irreversible around the transition temperature and increases the energy barrier which needs to be overcome through the PM-FM transition. Consequently the transition temperature on first cooling is found to be lower than on subsequent cycles characterizing the so-called ``virgin effect''. High-temperature annealing can restore the cycled sample to the high-temperature metastable state, which leads to the recovery of the virgin effect. A model is proposed to interpret the formation and recovery of the virgin effect.

\end{abstract}


\maketitle 

\begin{figure}         
 \includegraphics[trim =0mm 2mm 0mm 6mm, clip, scale=0.9]{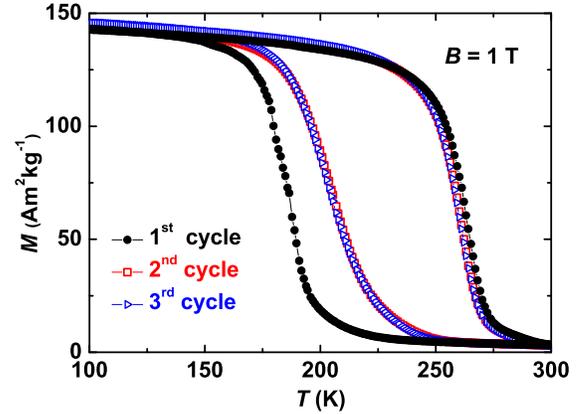}
\vspace{-0.8cm} 
 \caption{\label{VirginMT}(Color online) Temperature dependence of magnetization during the first, second and third thermal cycles of the as-prepared \BPChem{Mn\_{1.0}Fe\_{0.95}P\_{0.60}Si\_{0.40}} sample.}
 \vspace{-0.5cm}
\end{figure}

\BPChem{Fe\_{2}P}-type compounds, showing a strong spin-lattice coupling and a giant magnetocaloric effect, are of
great interest for scientists working on both fundamental research and technological applications. \cite{Fe2P, Tegus, Ge, DungAPL2011, DungAEM2011, CalculationFe2P, BCSitePreferenceFe2P} \BPChem{(Mn,Fe)\_{2}(P,Si)} compounds display a first-order paramagnetic-ferromagnetic (PM-FM) transition coupled with discontinuous changes in the lattice parameters.\cite{DungAPL2011, DungAEM2011, DungPRB2012} The character of the phase transition and the critical temperature can be easily tuned by balancing the Mn/Fe and P/Si ratios.\cite{MiaoPRB2014} Interestingly, the magnetoelastic transition in the \BPChem{(Mn,Fe)\_{2}(P,Si)} compounds shows a peculiar thermal-history dependence. As illustrated in Fig. \ref{VirginMT}, the as-prepared sample has a significantly lower phase transition temperature (\textsl{\BPChem{T\_{C}}}) upon first cooling than on second and subsequent cooling cycles. Since this behavior is only observed in as prepared or virgin samples it is termed the ``virgin effect''. Similar behavior has also been reported in \BPChem{MnAs}-based \cite{MnAsvirgin}, \BPChem{MnCoGe}-based \cite{Trungthesis}, \BPChem{(Mn,Fe)\_{2}(P,Ge)} \cite{Liu, Ekkes} and \BPChem{(Mn,Fe)\_{2}(P,Si,Ge)} \cite{Cam} compounds. \\
The difference in \textsl{\BPChem{T\_{C}}} between the first and second cooling processes of the as-prepared sample, hereafter referred to as $\Delta\textsl{\BPChem{T\_{Ci}}}$, is taken as a measure of how strong the virgin effect is. $\Delta\textsl{\BPChem{T\_{Ci}}}$ values of about 6 K and 13 K were observed in as-prepared \BPChem{MnAs}-based and \BPChem{MnCoGe}-based compounds, respectively. The virgin effect in these two compounds is due to the big volume change (about 2\% and 4\%) accompanying the magnetostructural transition.\cite{Trungthesis, MnAsvirgin} However, \BPChem{(Mn,Fe)\_{2}(P,Si)} has a stronger virgin effect with a $\Delta\textsl{\BPChem{T\_{Ci}}}$ of about 15 K, while the volume change at the magnetoelastic transition is less than 0.2\%. Zhang \textsl{et al}.\cite{Zhang} suggested that the virgin effect in \BPChem{(Mn,Fe)\_{2}(P,Si)} is due to the swapping of atomic positions during the first cooling of an as-prepared sample, although no experimental evidence was found. Based on temperature-dependent M{\"o}ssbauer experiments, Liu \textsl{et al}.\cite{Liu} proposed that the virgin effect observed in \BPChem{(Mn,Fe)\_{2}(P,Ge)} may originate from an additional irreversible structural change during the first cooling process of an as-prepared sample. H\"{o}glin \textsl{et al}.\cite{Viktor} also attributed the virgin effect in \BPChem{(Mn,Fe)\_{2}(P,Si)} to irreversible structural changes. In the present work we report evidence that supports the presence of an irreversible structural transformation as the cause of the virgin effect in \BPChem{(Mn,Fe)\_{2}(P,Si)}. We performed \textsl{in-situ} high-resolution neutron diffraction to monitor the changes in structural parameters, especially the interplanar spacing, during the first and second thermal cycle of an as-prepared \BPChem{(Mn,Fe)\_{2}(P,Si)} sample. A recovery of the virgin effect induced by thermal activation was observed experimentally.

The \BPChem{Mn\_{1.0}Fe\_{0.95}P\_{0.60}Si\_{0.40}} compound was prepared by ball-milling. The obtained fine powder was pressed into tablets and sealed in quartz ampoules. The samples were sintered at 1373 K for 2 h and then annealed at 1123 K for 20 h before being oven cooled to room temperature. The magnetic properties were characterized using a superconducting quantum interference device (SQUID) magnetometer (Quantum Design MPMS 5XL) in the reciprocating sample option (RSO) mode. \textsl{In-situ} neutron diffraction experiments were carried out on the time-of-flight high-resolution powder diffractometer (HRPD) at the ISIS pulsed neutron source facility, Rutherford Appleton Laboratory, UK. This instrument has a $\Delta d/d$ resolution of $\sim 4\times 10^{-4}$, which allows us to accurately study the changes in the interplanar spacings inside the sample through the phase transition. Neutron diffraction data were collected from three banks after thermal equilibrium at the following temperatures: 300, 185 ($\sim \mathit{T_{C}}$) and 120 K for the $1^{st}$ cooling; 300, 200 ($\sim \mathit{T_{C}}$) and 120 K for the $2^{nd}$ cooling. Nuclear and magnetic structure refinement of the neutron diffraction patterns was performed using the Rietveld method implemented in Fullprof \cite{Fullprof}.

\begin{figure}
 \centering            
  \includegraphics[trim = 0mm 0mm 0mm 0mm, clip, scale=0.85]{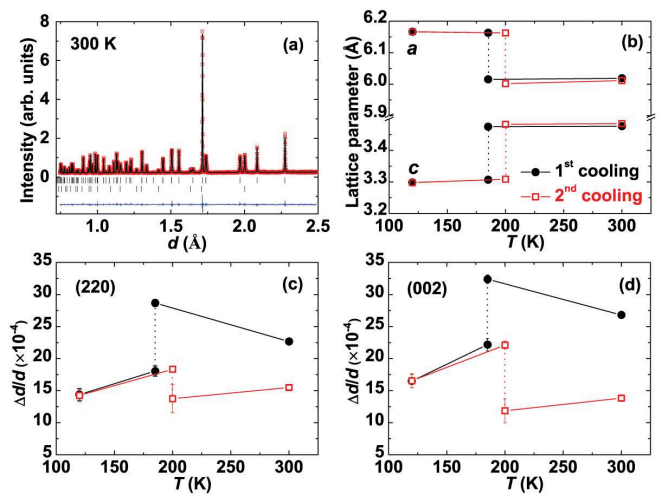}
	\vspace{-0.6cm} 
\caption{\label{Neutron}(Color online) (a) Observed and calculated neutron diffraction patterns at a $2\theta$ of $168.33^{\circ}$ for as-prepared \BPChem{Mn\_{1.0}Fe\_{0.95}P\_{0.60}Si\_{0.40}}. Vertical lines indicate the peak positions for the main phase (top) and the impurity phase \BPChem{(Mn,Fe)\_{3}Si} (bottom). Evolution of lattice parameters (b) and deviation of interplanar spacing of (220) atomic planes (c) and (002) atomic planes (d) during the first and second cooling of the as-prepared sample. The dotted line indicates the position of \textsl{\BPChem{T\_{C}}}.}
	\vspace{-0.5cm} 
\end{figure}

Fig. \ref{Neutron}a displays a neutron diffraction pattern of the as-prepared \BPChem{Mn\_{1.0}Fe\_{0.95}P\_{0.60}Si\_{0.40}} compound measured at 300 K, as an example. Good agreement between the experimental and calculated patterns was achieved assuming a hexagonal \BPChem{Fe\_{2}P}-type structure (space group $P\bar{6}2m$). Structural parameters calculated from the Rietveld refinement are summarized in Fig. \ref{Neutron}b-\ref{Neutron}d and Table \ref{Table}. As shown in Fig. \ref{Neutron}b and Table \ref{Table}, at 300 K the as-prepared sample has lattice parameters, atomic positions and occupancies very close to those after the first thermal cycle. The evolution of the lattice parameters also shows a similar behavior during the first two cycles, although a significantly lower \textsl{\BPChem{T\_{C}}} is observed for the first cooling.\\ 
\begin{table}
\centering
\caption{\label{Table} Structural parameters of \BPChem{Mn\_{1.0}Fe\_{0.95}P\_{0.60}Si\_{0.40}} at 300 K. Space group: $P\bar{6}2m$. Atomic positions: $3f$ (\BPChem{\textsl{x}\_1},0,0); $3g$ (\BPChem{\textsl{x}\_2},0,1/2); $2c$ (1/3,2/3,0) and $1b$ (0,0,1/2).}
\begin{tabular}{p{0.5cm} p{2.5cm} p{2.5cm} p{2.5cm}}
\hline
\hline
& Parameters & As-prepared state & After first cycle \\
\hline
3\textsl{f} & \BPChem{\textsl{x}\_1}	& 0.2538(2)	& 0.2538(2)\\
& \textsl{n}(Fe)/\textsl{n}(Mn) &	0.225/0.025(1) &	0.225/0.025(1)\\
3\textsl{g} & \BPChem{\textsl{x}\_{2}}	& 0.5902(4)	& 0.5895(4)\\
& \textsl{n}(Mn)/\textsl{n}(Fe)	& 0.238/0.012(1)	& 0.238/0.012(1)\\
2\textsl{c} & \textsl{n}(P)/\textsl{n}(Si)	& 0.094/0.073(4)	&0.094/0.073(4)\\
1\textsl{b} & \textsl{n}(P)/\textsl{n}(Si)	& 0.056/0.027(4)	& 0.056/0.027(4)\\
& \textsl{Rp}(\%)	& 3.43	& 3.63	\\
& \textsl{wRp}(\%)	& 4.18	& 4.33 \\
& \BPChem{{$\chi$}\^{2}} &	3.69 &	3.87\\
\hline
\hline
\end{tabular}
\vspace{-0.3cm}
\end{table}
It should be noted that the lattice parameters, atomic positions and occupancies obtained from the Rietveld refinement correspond to averages.\cite{DisorderNature, DisorderPRB, DisorderJAC} Local variations only contribute to the peak width. The local variations can be characterized by the variations in interplanar spacing, i.e., $\Delta d/d$. This quantity can be obtained using an anisotropic peak broadening model \cite{Strainmodel} to describe the peak shape of  our high-resolution neutron diffraction patterns. The temperature-dependent $\Delta d/d$ of (220) and (002) planes corresponds to variations within the basal \textsl{ab} plane and along the \textsl{c} axis, respectively (Fig. \ref{Neutron}c and \ref{Neutron}d). At 300 K the  as-prepared sample has much higher $\Delta d/d$ values for the (220) and (002) planes than after undergoing thermal cycling. The difference is more pronounced along the \textsl{c} axis. A drop in $\Delta d/d$ was observed around \textsl{\BPChem{T\_{C}}} (see Fig. \ref{Neutron}c and \ref{Neutron}d) during the first cooling. On the contrary, a reversible jump in $\Delta d/d$ around \textsl{\BPChem{T\_{C}}} appears during the second and subsequent cooling processes. This indicates that atoms in the as-prepared sample show a larger spread around their equilibrium crystallographic site, while the spatial variations become significantly smaller when the sample crosses the PM-FM transition for the first time. This additional atomic reconfiguration may increase the energy barrier to be overcome during the first PM-FM phase transition, which leads to a lower \textsl{\BPChem{T\_{C}}}. The subtle atomic reconfiguration is irreversible. As a result, a different \textsl{\BPChem{T\_{C}}} on cooling can only be observed when the as-prepared sample crosses the phase transition for the first time, being stabilized afterwards. The reversible jump in the $\Delta d/d$ around the \textsl{\BPChem{T\_{C}}} during the subsequent thermal cycles is due to the lattice distortion caused by the first-order magnetoelastic phase transition. Consequently, the same \textsl{\BPChem{T\_{C}}} is observed during the second and subsequent cooling processes.

The difference in atomic configuration between the as-prepared and cycled samples can be understood by considering the thermal history of the sample. As-prepared samples undergo a high-temperature sintering and annealing process where the high-temperature phase may be frozen and preserved at lower temperatures. However, the frozen high-temperature phase with a higher structural disorder is not stable at low temperatures and transforms to a more ordered stable phase when the sample crosses the PM-FM phase transition for the first time. This structural transition is irreversible at low temperatures since the activation energy cannot be reached, while high temperature annealing should provide enough energy to restore the initial high-temperature phase. As a consequence, the virgin effect may be recovered. To test this hypothesis, the as-prepared sample was first thermally cycled a few times using liquid nitrogen to eliminate the virgin effect. Subsequently, the cycled samples were re-annealed at different temperatures for 2 hours before being quenched into water at room temperature.

\begin{figure}
\centering            
  \includegraphics[trim = 0mm 0mm 0mm 0mm, clip, scale=0.75]{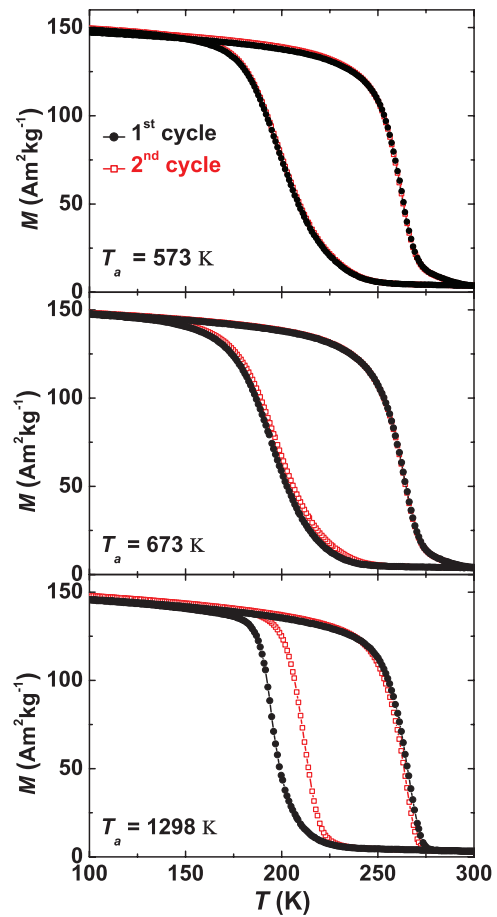}
		\vspace{-0.3cm} 
	\caption{\label{ReannealedMT}(Color online) Isofield magnetization curves measure at a field of 1 T during the first and second thermal cycles of the \BPChem{Mn\_{1.0}Fe\_{0.95}P\_{0.60}Si\_{0.40}} samples re-annealed at different temperatures.}
		\vspace{-0.2cm} 
\end{figure}

\begin{figure}
       \centering            
  \includegraphics[trim = 0mm 0mm 0mm 0mm, clip, scale=0.8]{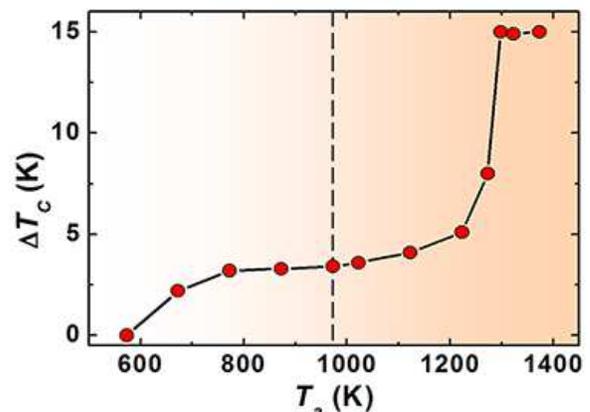}
			\vspace{-0.5cm} 
\caption{\label{DeltaTc}(Color online) $\Delta\textsl{\BPChem{T\_{C}}}$ as a function of re-annealing temperature. The solid lines are the guide to the eye.}
\vspace{-0.3cm} 
\end{figure}

The temperature-dependent magnetization of the re-annealed samples is shown in Fig. \ref{ReannealedMT}, from which the $\Delta\textsl{\BPChem{T\_{C}}}$ as a function of re-annealing temperature (\textsl{\BPChem{T\_{a}}}) can be obtained (see Fig. \ref{DeltaTc}). No detectable shift in \textsl{\BPChem{T\_{C}}} between the first and second cooling can be observed when \textsl{\BPChem{T\_{a}}} is lower than 573 K. At 673 K, a $\Delta\textsl{\BPChem{T\_{C}}}$ of about\linebreak 2 K appears, indicating the start of the structural recovery process. $\Delta\textsl{\BPChem{T\_{C}}}$ increases slowly until \textsl{\BPChem{T\_{a}}} $\simeq$ 973 K. Above this temperature the recovery process accelerates with increasing re-annealing temperature. $\Delta\textsl{\BPChem{T\_{C}}}$ is fully recovered at \textsl{\BPChem{T\_{a}}} from 1298 K up to the original sintering temperature of 1373 K. Another feature that can be seen from Fig. \ref{ReannealedMT} is that the PM-FM transition of the 1298 K re-annealed sample is much sharper than the as-prepared and the low-\textsl{\BPChem{T\_{a}}} re-annealed ones. This is due to better compositional homogeneity obtained after the high-temperature re-annealing process.\\
\begin{figure}
       \centering            
  \includegraphics[trim = 0mm 0mm 0mm 0mm, clip, scale=0.85]{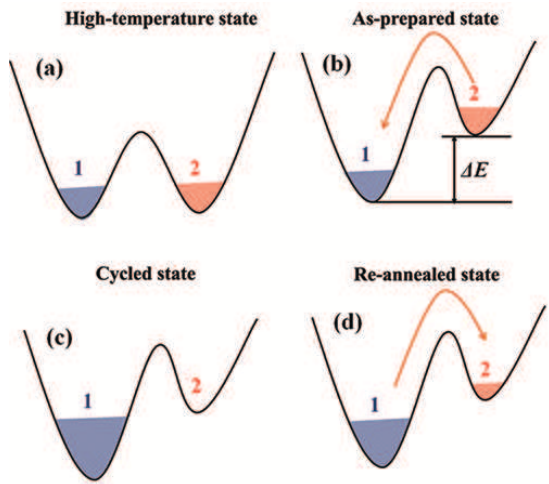}
			\vspace{-0.3cm} 
\caption{\label{Energy}(Color online) Schematic diagram of free-energy profiles for \BPChem{(Mn,Fe)\_{2}(P,Si)} during thermal cycling.}
	\vspace{-0.3cm} 
\end{figure}
As discussed above, the as-prepared sample has a higher structural disorder that results in an additional irreversible structural transition during the first cooling process. High temperature annealing can restore the metastable phase, and consequently recover the virgin effect. Based on these results, we propose a model to describe the virgin effect in \BPChem{(Mn,Fe)\_{2}(P,Si)} compounds. The high thermal activation energies accessible at elevated temperatures allow atoms to occupy positions which deviate from the crystallographic site described in Table \ref{Table}. There may be two or multiple equivalent energy-minimum states in the system at high temperature, which correspond to different atomic coordinates in the structure. Currently, with the experimental error at hand we cannot determine how many sites are involved. In our model, we assume that there are two energetically degenerated states in the system, as illustrated in Fig. {\ref{Energy}}a. The high-temperature atomic configuration may be preserved in the as-prepared sample after high-temperature sintering and annealing. However, the free-energy profile is different at low temperature and thus the states 1 and 2 are not equivalent any more. As the sample is cooled down an energy barrier $\Delta\textsl{\BPChem{E}}$ appears between the stable state (described by state 1 in Fig. \ref{Energy}b) and metastable state (illustrated by state 2 Fig. \ref{Energy}b). The virgin effect depends on the relative populations of the two states, which can be described as
\begin{equation}
\Delta {T_{C}} = \frac{2 \Delta {T_{Ci}}}{1+e^{\Delta E/RT}}
\end{equation}
where $\Delta\textsl{\BPChem{T\_{C}}}$ and $\Delta\textsl{\BPChem{T\_{Ci}}}$ are the difference in $\textsl{\BPChem{T\_{C}}}$ between the first and second cooling for the re-annealed and as-prepared samples, respectively. The $\Delta\textsl{\BPChem{E}}$ is the effective energy barrier between the states 1 and 2, related to the populations $\textsl{\BPChem{N\_{1}}}$ and $\textsl{\BPChem{N\_{2}}}$ of states 1 and 2, respectively, by the Boltzmann factor $N_{2}/N_{1} = e^{-\Delta E/RT}$. $\textsl{R}$ is the gas constant.

During the first cooling of the as-prepared sample, the frozen high-temperature metastable phase transforms to a stable phase through the atomic reconfiguration caused by the magnetoelastic phase transition, as illustrated by Fig. \ref{Energy}b. After the first thermal cycle only state 1 is occupied (see Fig. \ref{Energy}c), i.e., \textsl{\BPChem{N\_{1}}} = 1 and \textsl{\BPChem{N\_{2}}} = 0, which results in a more ordered state as observed by neutron diffraction. The structural transition is irreversible at low temperatures because of the low thermal energy and the higher energy barrier. Therefore $\Delta\textsl{\BPChem{T\_{C}}}$ is always 0 after the first thermal cycle. When the cycled sample is re-annealed at intermediate temperatures (between 673 and 973 K as indicated in Fig. \ref{DeltaTc}), thermal energy is enough to start overcoming the energy barrier between states 1 and 2. As a result, \textsl{\BPChem{N\_{2}}} increases gradually and so does $\Delta\textsl{\BPChem{T\_{C}}}$. For higher \textsl{\BPChem{T\_{a}}}, the energy barrier between states 1 and 2 decreases, which leads to a rapid increase of \textsl{\BPChem{N\_{2}}} and $\Delta\textsl{\BPChem{T\_{C}}}$. As \textsl{\BPChem{T\_{a}}} goes above 1298 K, the initial high-temperature scenario with equivalent energy-minimum states (Fig. \ref{Energy}a) is restored and the virgin effect is fully recovered.

The virgin effect of \BPChem{MnCoGe}-based and \BPChem{MnAs}-based compounds may also originate from a metastable state in the as-prepared sample. As-prepared \BPChem{MnCoGe}-based compounds have a hexagonal structure, which is formed at room temperature by quenching the sample from high temperature.\cite{MnCoGediagram} The frozen high-temperature phase may have higher disorder giving rise to the virgin effect in a similar way as in \BPChem{Fe\_{2}P}-based compounds. \BPChem{MnAs}-based compounds crystallize in a re-entrant orthorhombic structure, while showing  the same hexagonal structure at both high and low temperatures.\cite{MnAsdiagram} The structural transition between the high-temperature hexagonal and the re-entrant orthorhombic phase may cause structural distortions in the as-prepared orthorhombic structure, which increases the energy barrier of the magnetostructural transition between re-entrant orthorhombic and low-temperature hexagonal phases. During the subsequent thermal cycles, the distorted orthorhombic phase does not appear again since the sample does not enter the high-temperature hexagonal region. As a result, a lower \textsl{\BPChem{T\_{C}}} is observed only when the as-prepared \BPChem{MnAs}-based compound undergoes the orthorhombic to low-temperature hexagonal transition for the first time.

In conclusion, the unusual thermal-history dependent phase transition behavior in as-prepared \BPChem{(Mn,Fe)\_{2}(P,Si)} compounds, called the ``virgin effect'' was studied.  The virgin effect is found to be related to a metastable state preserved in as-prepared samples after high-temperature annealing where large deviations of atomic positions from the crystallographic sites are observed. The frozen metastable phase transforms irreversibly to the stable low-temperature phase during the paramagnetic-ferromagnetic transition upon first cooling and is afterwards stabilized. Due to this additional transformation the energy barrier which needs to be overcome at the first cooling transition is higher and a lower \textsl{\BPChem{T\_{C}}} is observed in relation to subsequent cooling processes. High temperature annealing can restore the high-temperature metastable state, which leads to the recovery of the ``virgin effect''.
 
The authors would like to thank Anton Lefering and Bert Zwart for their help with the sample preparation. This work is part of the Industrial Partnership Program of the Dutch Foundation for Fundamental Research on Matter (FOM), and co-financed by BASF New Business.
%
\end{document}